\documentclass[conference]{IEEEtran}
\usepackage{epsfig}
\usepackage{times}
\usepackage{float}
\usepackage{afterpage}
\usepackage{amsmath}
\usepackage{amstext}
\usepackage{amssymb,bm}
\usepackage{latexsym}
\usepackage{color}
\usepackage{graphicx}
\usepackage{amsmath}
\usepackage{amsthm}
\usepackage{graphicx}
\usepackage[center]{caption}
\usepackage{pstricks}
\usepackage{caption}
\usepackage{subcaption}
\usepackage{booktabs}
\usepackage{multicol}
\usepackage{lipsum}% dummy text
 \usepackage[T1]{fontenc}
\usepackage{epstopdf}

\allowdisplaybreaks

\newtheorem{thm}{Theorem}%[section]

\theoremstyle{definition}

\providecommand{\definitionname}{Definition}

\begin{document}

\title{On Caching with More Users than Files}

\author{\IEEEauthorblockN{Kai~Wan}\IEEEauthorblockA{Laboratoire des Signaux et Système (L2S)\\
CentraleSupélec-CNRS-Université Paris-Sud\\
Gif-sur-Yvette, France\\
Email: kai.wan@u-psud.fr}\and \IEEEauthorblockN{Daniela~Tuninetti}\IEEEauthorblockA{University of Illinois at Chicago\\
Chicago, IL 60607, USA\\
Email: danielat@uic.edu}\and \IEEEauthorblockN{Pablo~Piantanida}\IEEEauthorblockA{Laboratoire des Signaux et Système (L2S)\\
CentraleSupélec-CNRS-Université Paris-Sud\\
Gif-sur-Yvette, France\\
Email: pablo.piantanida@centralesupelec.fr}}
%\author{\IEEEauthorblockN{Kai~Wan}\IEEEauthorblockA{Laboratoire des Signaux et Système (L2S)\\
%Centrale-CNRS-Université Paris-Sud\\
%Gif-sur-Yvette, France\\
%Email: kai.wan@u-psud.fr}\and \IEEEauthorblockN{Daniela~Tuninetti}\IEEEauthorblockA{University of Illinois at Chicago\\
%Chicago, IL 60607, USA\\
%Email: danielat@uic.edu}\and \IEEEauthorblockN{Pablo~Piantanida}\IEEEauthorblockA{Laboratoire des Signaux et Système (L2S)\\
%Centrale-CNRS-Université Paris-Sud\\
%Gif-sur-Yvette, France\\
%Email: pablo.piantanida@centralesupelec.fr}}

\maketitle

\begin{abstract}
Caching appears to be an efficient way to reduce peak hour network traffic congestion
by storing some content at the user's cache without knowledge of later
demands. Recently, Maddah-Ali and Niesen proposed a two-phase, placement
and delivery phase, coded caching strategy for centralized systems
(where coordination among users is possible in the placement phase),
and for decentralized systems. %(with random content placement).
This paper investigates the same setup under the further assumption that the number of users is larger than the number of files. 
By using the same uncoded placement strategy of Maddah-Ali and Niesen, 
a novel coded delivery strategy is proposed
to profit from the multicasting opportunities that arise because
a file may be demanded by multiple users. 
The proposed delivery method is proved to be optimal under the constraint of uncoded placement for centralized systems with two files; moreover it is shown to outperform known caching strategies
for both centralized and decentralized systems.
% which is better than their schemes.
\end{abstract}

\IEEEpeerreviewmaketitle{}

\section{Introduction}
\label{sec:intro}
%Due to the variability of network traffic in a broadcasting system,
%we can divide the whole time into peak hours and off peak hours. Each
%user can be equipped with a memory to store some contents without
%knowledge of later demands during the off peak hours such that the
%load over a shared link during the peak hours can be reduced. This
%is the core of a widely used technique in practice, named caching. 
Caching is a popular method to smooth out network traffic in broadcasting systems,
where some content is cached into the user's memory during off peak hours
in the hope that the pre-stored content will be required by the user during peak hours and  thus, reducing the number of broadcast transmissions from the server to the users.

\subsubsection*{System model}
In this paper, we study a system with  $N$ files available at a server that is connected to $K$ users; each user has a cache of size $M$ to store files; users are connected to the server via a shared error-free link. The caching procedure assumes two phases. (1) Placement phase: where users store (coded or uncoded) pieces of the files within their cache without knowledge of later demands.  When the file pieces are not network coded we say that  the placement phase is {\it uncoded}, otherwise that it is {\it coded}. (2) Delivery phase: where each user demands a specific file and, based on the users' demands and cache content,  the server broadcasts packets %over an error-free link  to all users 
so that each user can recover the demanded file.  The objective of the system designer is to provide a two-phase scheme so that the number of transmitted packets, or load, in the delivery phase for the worst-case demands is minimized.

\subsubsection*{Coordinated cache placement}
Maddah-Ali and Niesen proposed~\cite{dvbt2fundamental} a coded caching scheme that utilizes an uncoded combinatorial cache construction in the placement phase and a linear network code in the delivery phase, where users store contents in a coordinated manner.  The worst-case load of the  Maddah-Ali and Niesen scheme (refer to as MNS) was shown to be no larger than
$K\left(1-\frac{M}{N}\right)\min \left\{ \frac{1}{1+K\frac{M}{N} },\frac{N}{K}\right\}$,
which has the additional global caching gain $\frac{1}{1+K M/N}$ 
compared to the conventional uncoded caching scheme. MNS was shown to be  optimal~\cite{ontheoptimality} 
under the constraint of uncoded cache placement and $N\geq K$,  and order optimal~\cite{dvbt2fundamental}  to within a factor of $12$ of the cut-set outer bound. The authors in~\cite{smallbufferusers} showed that a scheme based on coded cache placement, originally proposed in~\cite{dvbt2fundamental} for $N=2$, is optimal when $N\leq K$ and $MK\leq1$ while providing a load of $N(1-M)$ which coincides with the cut-set outer bound. Recently, reference~\cite{kuserstwofiles} studied the case $N=2$ and $M\leq\frac{K-1}{K}$, and proposed a scheme with coded cache placement yielding a lower load than MNS.

\subsubsection*{Un-coordinated cache placement} 
The previously mentioned works 
%are based on coordinated cache placement, which needs a centralized entity to decide on the content placement. This 
assumed that the $K$ connected users are the same during both phases.  However, this may not always be the case  in practice (e.g. due to user mobility) where a user may be connected to one server 
during his placement phase but to a different one during his delivery phase. In this decentralized scenario, each server must carry out independently the two phases of caching and thus, the coordination (among users) during  the placement phase is not possible. In~\cite{decentralizedcoded}, Maddah-Ali and Niesen proposed that each user fills its cache randomly and independently of the others. During the delivery phase, the bits of $N$ files are organized into sub-files depending  on which users know, each of which is delivered by using the delivery strategy  in~\cite{dvbt2fundamental} for centralized systems. The corresponding load was shown to be $K\left(1-\frac{M}{N}\right)\min \left\{ \frac{N}{KM}(1-(1-\frac{M}{N})^{K}),\frac{N}{K}\right\}$, where the factor $\frac{N}{KM}\left[1-(1-\frac{M}{N})^{K}\right]$ represents an additional global caching gain compared to the conventional uncoded caching.

A delivery phase with load equal to the fractional local chromatic number  (described in~\cite{indexcodingrandom}) of the directed graph formed by the users' demands and caches was shown in~\cite{groupcastindexcoding, detailsgroupe} for centralized and decentralized scenarios, respectively. Since the computation of the  fractional local chromatic number is NP-hard, the authors in~\cite{simplifiedgroupeindexcoding, grasp} proposed approximate algorithms to simplify computations.

\subsubsection*{Our contribution}
In~\cite{ontheoptimality}, we showed that for  $N\geq K$ and under the constraint of uncoded cache placement, MNS is optimal. %This proof was based on the identification of the caching problem as being an instance of the index coding problem. 
In this work, motivated by practical considerations (e.g., 
a server has several popular music or video files that are widely demanded by different users), we study the case $N<K$ where same sub-files may be demanded by multiple users. It is worthing noting that MNS cannot be used to multicast files since it considers each sub-file demanded by each user as a district sub-file. With the goal of multicasting messages, we design a delivery phase for the case of $N<K$ that is applicative to both centralized and decentralized scenarios. The proposed delivery method is shown to achieve the optimal load under the constraint of uncoded placement for centralized systems with two files and to outperform known caching strategies for both centralized and decentralized scenarios. 

\subsubsection*{Paper Outline}
The rest of the paper is organized as follows. 
Section~\ref{sec:model} presents the system model.
Section~\ref{sec:newach} introduces the main results.
Section~\ref{sec:num} %shows the simulation results of our schemes 
compares by numerical results the proposed scheme to existing ones.
Finally, Section~\ref{sec:conclusions} presents summary and discussion while some technical proofs are relegated to the Appendix.

\subsubsection*{Notations}
Calligraphic symbols denotes sets;
$|\cdot|$ is used to represent the cardinality of a set or the length of a file;
we denote $[1:K]:=\left\{ 1,2,...,K\right\}$ and $\mathcal{A\setminus B}:=\left\{ x\in\mathcal{A}|x\notin\mathcal{B}\right\}$;
$\oplus$ represents the bit-wise XOR operation,
and $\binom{K}{t}$ is the binomial coefficient.

\section{System Model and Problem Statement}
\label{sec:model}
Consider a broadcasting caching system that consists of a center server with $N$ files, denoted by $(F_{1},F_{2},\dots,F_{N})$, and $K$ users connected to it through an error-free link.  Each file has $F\gg 1$ bits. Here we assume $N<K$ and that each file is requested by each user with identical probability. 

During the placement phase, user $i\in[1:K]$ stores content from $N$ files in his cache of size $MF$ bits without knowledge of later demands, where $M\in[0,N]$. We denote the content in the cache of user $i$ by $Z_{i}$; we also let $\mathbf{Z}:=(Z_{1},\dots,Z_{K})$. Centralized systems allow for coordination among users in the placement phase, while decentralized systems do not. In the delivery phase, each user demands one file and the demand vector $\mathbf{d}:=(d_{1},d_{2},\dots,d_{K})$ is revealed to the server, where $d_{i}\in[1:N]$ is the file demanded by user $i\in[1:K]$.

Given $(\mathbf{Z},\mathbf{d})$, the server broadcasts a message $X_{\mathbf{d},\mathbf{Z}}$ with normalized length (by the file size $F$) $R(\mathbf{d},M)$. It is  required that user $i\in[1:K]$ recovers his desired file $F_{d_{i}}$ from $X_{\mathbf{d},\mathbf{Z}}$ and $Z_{i}$ with high probability. The objective is to minimize the worst-case network load:
$
R(M)=\mathrm{min}\thinspace\thinspace \underset{\mathbf{d}}{\mathrm{max}}\thinspace\thinspace R(\mathbf{d},M).
$
%is the worst-case normalized load for a caching scheme. In the remainder
%of this paper for the simplicity, load of a scheme means the normalized
%load of this scheme.

\section{Main Results}
\label{sec:newach}
%Our main results are as follows. 
%
%For centralized systems, 
We propose a caching scheme that attains the following memory-load tradeoffs for centralized systems.
\begin{thm}[Centralized]% system
\label{thm1 uncoded placement load} 
For centralized systems, the lower convex envelope of $R_\mathrm{p}(M)$ is achievable with $t\in[0:K]$ and
\begin{align}
R_{\mathrm{c}}(M) & =\begin{cases}
N(1-M), & M=\frac{1}{K},\\
R_{\mathrm{co}}(M), & M=t\frac{N}{K}, 0\leq M< M_\text{th},\\
\frac{K(1-\frac{M}{N})}{1+K\frac{M}{N}}, & M=t\frac{N}{K}, M_\text{th}\leq M\leq N, %\thinspace\mathrm{and}\thinspace
\end{cases}\label{eq:Rp(M)}\\
R_{\mathrm{co}}(M) & =N-M-\frac{M(N-1)K(N-M)}{N^{2}(K-1)},\label{eq:rpofM1}\\
M_{\mathrm{th}} & :=N\frac{NK-2N+1-\sqrt{f(N,K)}}{2K(N-1)},\label{eq:Mth}\\
f(N,K) & :=(NK-2N+1)^{2}-4(N-1)(K-N)(K-1).\label{eq:f(N,K)}
\end{align}
\end{thm}

The same idea applied to decentralized systems attains the following memory-load tradeoff.
\begin{thm}[Decentralized]% system
\label{thm2 decentralized load} 
For decentralized systems, the lower convex envelope of $R_\mathrm{d}(M)$ is achievable with $M\in[0,N]$, $q :=M/N$ and 
\begin{align}
 & R_\mathrm{d}(M)=N(1-q)\mathrm{C}(t_\mathrm{th},K-1,q)-(N-1)q(q-1)\nonumber \\
 & \mathrm{C}(t_\mathrm{th}-1,K-2,q)+\frac{1-q}{q}(1-\mathrm{C}(t_\mathrm{th}+1,K,q)),
\label{eq:CDF forme rd(M)}
\\
 & \mathrm{C}(x,y,q) :=\sum\limits _{i=0}^{x}\binom{y}{i}q^{i}\left(1-q\right)^{y-i}.\label{eq:C(x,y,q)}
\end{align}
\end{thm}

We next derive an outer bound under the constraint of uncoded cache placement and $K>N$ and prove the optimality of the proposed achievable scheme for $N=2$.
\begin{thm}[Optimality for $N=2$]
\label{thm3 uncoded placement optimality} 
The minimal load under the constraint of uncoded cache placement and $K>N=2$ for the aforementioned centralized systems, is $R_{\mathrm{co}}(M)$ in~\eqref{eq:rpofM1} and is achieved by the proposed scheme.
\end{thm}
The rest of the Section is firstly devoted to the proof of Theorem~\ref{thm1 uncoded placement load} and Theorem~\ref{thm2 decentralized load}. The main idea %of the proposed delivery coding 
is to consider the multicasting opportunities that arise for the case of $N<K$. Due to space limitation, 
%we only present an outline of 
the proof of Theorem~\ref{thm3 uncoded placement optimality} is only outlined.

%We divided all the sub-files into
%several groups according to users' demands and then for each group
%we use an individual linear code. At last, a linear code is used for
%all the codes from last step. 

\subsection{Proof of Theorem~\ref{thm1 uncoded placement load}}
\label{sec:newach:centralized}

We start by describing our scheme and computing the load %the algorithm
for $M=t\frac{N}{K}$, where $t\in[0:K]$. 
The complete memory-load tradeoff is obtained as the lower convex envelope
of the derived points, which can be achieved by memory sharing.

\subsubsection*{Placement Phase}
The cache placement phase is as in the MNS. %the same as~\cite{dvbt2fundamental}.
Each file is split into $\binom{K}{t}$ non-overlapping sub-files of
identical size given by $\frac{F}{\binom{K}{t}}$, where $t=K\frac{M}{N}\in[0:K]$.
Each sub-file of $F_{i}$ is denoted by $F_{i,\mathcal{W}}$ where $\mathcal{W}\subseteq[1:K]$
such that $|\mathcal{W}|=t$. User $j\in[1:K]$ stores $F_{i,\mathcal{W}}$ for all $i\in[1:N]$ in his
cache if and only if $j\in\mathcal{W}$.

\subsubsection*{Delivery Phase}
The delivery phase is divided into two steps. 
We consider the worst case demand where each file is demanded by at least one user. Let $\mathcal{G}_{i}$ be the set of users
who demand file $F_{i}$, for $i\in [1:N]$.

\textbf{Step 1:}
We divide the sub-files of $F_{i}$ into several groups indicated as
$\mathcal{O}_{i,J} :=\{F_{i,\mathcal{W}}:\mathcal{W}\setminus\mathcal{G}_{i}=\mathcal{J}\}$,
where $\mathcal{J}\subseteq[1:K]\setminus\mathcal{G}_{i}$ and $\textrm{max}\{0,t-\mathcal{G}_{i}\}\leq |\mathcal{J}|\leq t$.
There are $\binom{|\mathcal{G}_{i}|}{t-|\mathcal{J}|}$ sub-files in $\mathcal{O}_{i,J}$. 
Each user in $\mathcal{G}_{i}$ wants to recover all the sub-files in $\mathcal{O}_{i,J}$ and knows $\binom{|\mathcal{G}_{i}|-1}{t-|\mathcal{J}|-1}$ of them. Note that when $|\mathcal{J}|=t$, we assume  $\binom{|\mathcal{G}_{i}|-1}{t-|\mathcal{J}|-1}=0$. The authors in~\cite{decentralizedcoded} showed that this kind of problem can be solved by using $m-d$ random linear combinations of all the $m$ bits, where $m$ and $d$ are number of bits to encode and minimum number of bits known at each decoder. Since $m$ and $d$ tend to infinite, the $m-d$ random linear combinations are linearly independent with high probability, thus each decoder can recover all the $m$ bits with high probability. Hence, in order to delivery all the sub-files in $\mathcal{O}_{i,J}$ to the users in $\mathcal{G}_{i}$, we can use 
$\ensuremath{\left[\binom{|\mathcal{G}_{i}|}{t-|\mathcal{J}|}-\binom{|\mathcal{G}_{i}|-1}{t-|\mathcal{J}|-1}\right]\frac{F}{\binom{K}{t}}}$ random linear combinations of all the $\binom{|\mathcal{G}_{i}|}{t-|\mathcal{J}|}\frac{F}{\binom{K}{t}}$ bits in $\mathcal{O}_{i,J}$. 
We define $\mathcal{C}_{i,J}$ as the code for $\mathcal{O}_{i,J}$. 
With the \emph{Pascal's triangle}
%%, using Lemma \ref{lem:An-achievable-generator} and
\begin{align}
\binom{|\mathcal{G}_{i}|}{t-|\mathcal{J}|}-\binom{|\mathcal{G}_{i}|-1}{t-|\mathcal{J}|-1}=\binom{|\mathcal{G}_{i}|-1}{t-|\mathcal{J}|},
\label{eq:pascal}
\end{align} 
it can be seen easily that $\mathcal{C}_{i,J}$ has
$\binom{|\mathcal{G}_{i}|-1}{t-|\mathcal{J}|}\frac{F}{\binom{K}{t}}$ bits. Note that when $|\mathcal{J}|=t-\mathcal{G}_{i}$, the right side of~\eqref{eq:pascal} is $0$. Let $v_{i,t}:=\textrm{max}\{0,t-\mathcal{G}_{i}+1\}.$
As a consequence, for each $\mathcal{J}$ where $\mathcal{J}\subseteq[1:K]\setminus\mathcal{G}_{i}$
and $v_{i,t}\leq |\mathcal{J}|\leq t$, we use random linear combinations as described above %use lemma \ref{lem:An-achievable-generator}
to encode $\mathcal{O}_{i,J}$. 
We define $\mathcal{C}_{i}$ as the set of $\mathcal{C}_{i,J}$ for all $\mathcal{J}\subseteq[1:K]\setminus\mathcal{G}_{i}$ and $v_{i,t}\leq |\mathcal{J}|\leq t$. 
The number of bits in $\mathcal{C}_{i}$ is equal to (see Appendix)
\begin{align}
\sum_{\mathcal{J}\subseteq[1:K]\setminus\mathcal{G}_{i}:v_{i,t}\leq |\mathcal{J}|\leq t}\binom{|\mathcal{G}_{i}|-1}{t-|\mathcal{J}|}\frac{F}{\binom{K}{t}}=\binom{K-1}{t}\frac{F}{\binom{K}{t}}.
\label{eq:app 1}
\end{align}
%The proof of~\eqref{eq:app 1} can be found in . 

Let $\mathcal{C}_\text{step1}$ denote set of bits in $\mathcal{C}_{i}$ for all $i\in[1:N]$, where $\mathcal{C}_\text{step1}$ has $N\binom{K-1}{t}\frac{F}{\binom{K}{t}}$ bits.
If the server transmits $\mathcal{C}_\text{step1}$, each user would be able to recover
his desired file with very high probability. 
%that Moreover, we can still compress it. 
However, by doing so we have some redundancy left,
which motivates the next step. 

\textbf{Step 2:}
For file $F_{i}$ and user $j\notin\mathcal{G}_{i}$,
user $j$ knows some bits in $\mathcal{C}_{i}$. More precisely,
user $j$ knows $\mathcal{C}_{i,J}$ if $j\in\mathcal{J}$ and hence, the number of bits in $\mathcal{C}_{i}$ known by $j$ is (see  Appendix)
\begin{align}
\sum_{\mathcal{J}\subseteq[1:K]\setminus\mathcal{G}_{i}:v_{i,t}\leq |\mathcal{J}|\leq t,j\in\mathcal{J}}\binom{|\mathcal{G}_{i}|-1}{t-|\mathcal{J}|}\frac{F}{\binom{K}{t}}=\binom{K-2}{t-1}\frac{F}{\binom{K}{t}}.
\label{eq:app2}
\end{align}
%The proof of~\eqref{eq:app2} can be found in Appendix. 

Considering $\mathcal{C}_{i}$ for all $i$ such that user $j\notin\mathcal{G}_{i}$, 
the total number of bits known by $j$
is $(N-1)\binom{K-2}{t-1}\frac{F}{\binom{K}{t}}$. We can use
 $\left[N\binom{K-1}{t}-(N-1)\binom{K-2}{t-1}\right]\frac{F}{\binom{K}{t}}$ random linear combinations to encode $\mathcal{C}_\text{step1}$.
%can be used such that 
As a result by letting  $t=\frac{KM}{N}\in[0:K]$, the load of our scheme is $R_\text{co}(M)$ in~\eqref{eq:rpofM1}.

Compared to the delivery method in the MNS, whose
%~\cite{dvbt2fundamental} worst-case 
load is $K(1-\frac{M}{N})\mathrm{min}\left\{ \frac{1}{1+K\frac{M}{N}},\frac{N}{K}\right\} $, it can be shown that for $0\leq M\leq N$,
\begin{align*}
R_\text{co}(M)\leq K(1-\frac{M}{N})\frac{N}{K}=N-M.
\end{align*}
We can also find that if $0\leq M< M_\text{th}$, 
\begin{align*}
R_\text{co}(M)&<K(1-\frac{M}{N})\frac{1}{1+K\frac{M}{N}},
\end{align*}
and that if $M_\text{th}\leq M\leq N$,
\begin{align*}
R_\text{co}(M)&\geq K(1-\frac{M}{N})\frac{1}{1+K\frac{M}{N}},
\end{align*}
where the threshold $M_\text{th}$ was given in~\eqref{eq:Mth}.
%So we can have a novel
%delivery scheme by choosing the proposed two-step method or the delivery
%method in~\cite{dvbt2fundamental} where the choice depends on whether
%$M$ is less than $M_\text{th}$ or not. 

Finally, by memory sharing the caching scheme in~\cite{smallbufferusers} (which is optimal for $M=1/K$) together with the above proposed scheme, 
we have the load is no larger than the lower convex envelope of $R_\text{p}(M)$ described in~\eqref{eq:Rp(M)}. \hfill\qed

\subsubsection*{Example} % for centralized system

\begin{figure}%[tbh]
\centering{}
\includegraphics[scale=0.4]{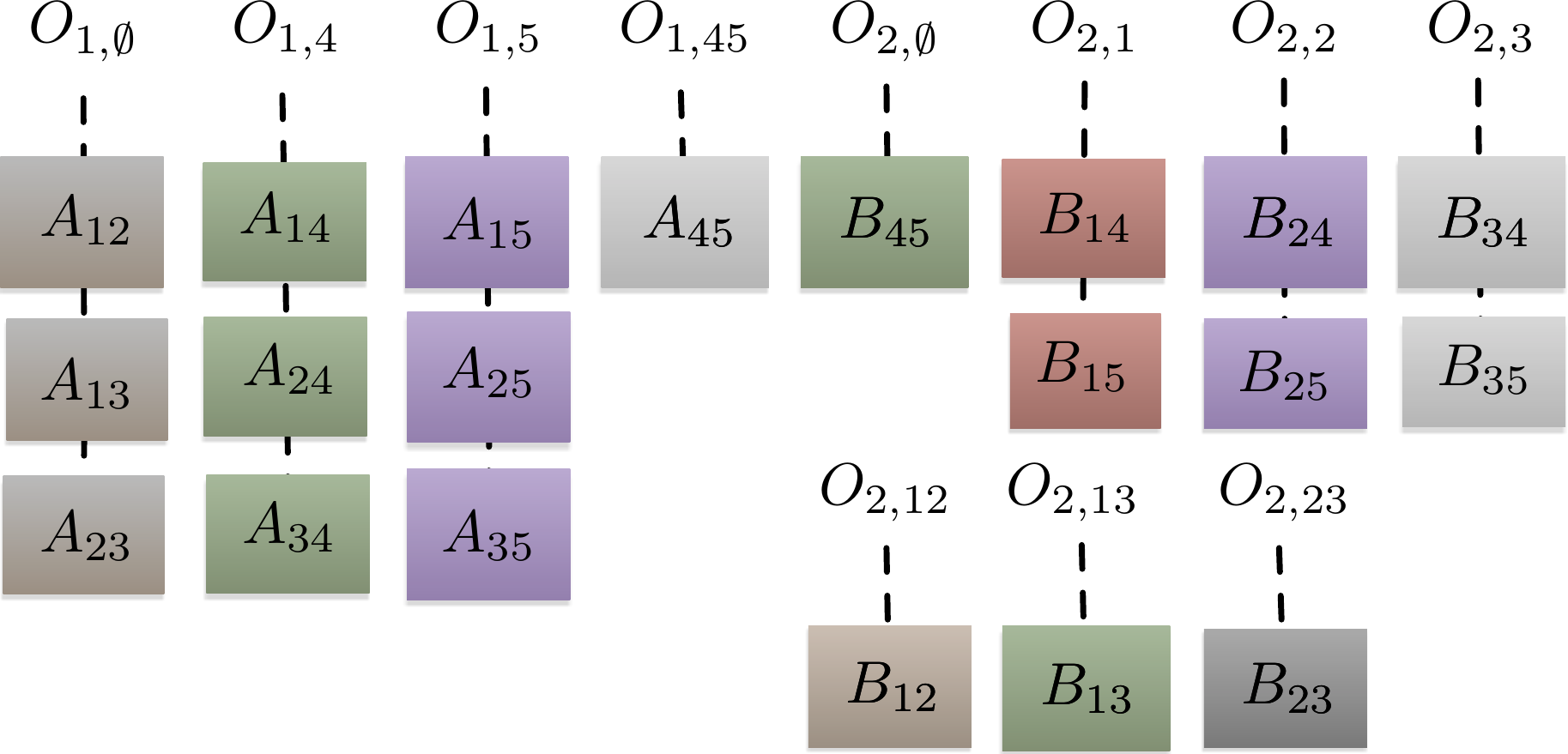}
\caption{Groups of sub-files for the example $N=2$, $K=5$, $M=4/5$ and $\mathbf{d}=(1,1,1,2,2)$.}
\label{fig:excent}
\end{figure}

In order to clarify the steps of the proposed scheme,  we analyse here in detail
the case $N=2$, $K=5$, $M=4/5$, and $F_{1}=A$, $F_{2}=B$.
With these parameters we have $t=\frac{KM}{N}=2$
and we therefore split each of the two files $A$ and $B$
into $\binom{K}{t}=10$ non-overlapping sub-files of size equal to
$\frac{F}{10}$. For simplicity
in the following we omit the braces when we indicate sets, i.e., $A_{12}$
represents $A_{\{12\}}$. 

In the placement phase we set $Z_{j}=\{F_{i,\mathcal{W}}:j\in\mathcal{W},|\mathcal{W}|=t\}$ for $i=\{1,2\}$, e.g., $Z_{1}=\{A_{12},A_{13},A_{14},A_{15},B_{12},B_{13},B_{14},B_{15}\}$. In the delivery phase, since $M_\text{th}%=N\frac{NK-2N+1-\sqrt{f(N,K)}}{2K(N-1)}
=1.2>M = 4/5=0.8$, 
the novel proposed two-step method is used. 
We consider the worse-case demand vector
% $\mathbf{d}=(1,1,1,2,2)$,
%%and treat each sub-file as a unit. 
%and thus 
$\mathcal{G}_{1}=\{1,2,3\}$ and $\mathcal{G}_{2}=\{4,5\}$.

In step~1, we divide the sub-files of $A$ into several groups, 
$\mathcal{O}_{1,J}=\{F_{1,\mathcal{W}}:\mathcal{W}\setminus\mathcal{G}_{1}=\mathcal{J}\}$,
where $\mathcal{J}\subseteq\{4,5\}$
and $|\mathcal{J}|\leq2$. 
Similarly, $\mathcal{O}_{2,J}=\{F_{2,\mathcal{W}}:\mathcal{W}\setminus\mathcal{G}_{2}=\mathcal{J}\}$,
where $\mathcal{J}\subseteq\{1,2,3\}$
and $|\mathcal{J}|\leq2$. 
The groups can be seen in Fig.~\ref{fig:excent}, identified by different colors.
For each group $\mathcal{O}_{i,J}$,
each user in $\mathcal{G}_{i}$ wants to recover all the $\binom{|\mathcal{G}_{i}|}{t-|\mathcal{J}|}$
sub-files in this group and knows $\binom{|\mathcal{G}_{i}|-1}{t-|\mathcal{J}|-1}$
of them. 
For instance, for $\mathcal{O}_{1,4}=\{A_{14},A_{24},A_{34}\}$,
each of the users in $\mathcal{G}_{1}=\{1,2,3\}$ wants to recover $\mathcal{O}_{1,4}$ whose length is $\frac{3F}{10}$,
while user~$1$ knows $A_{14}$, user~$2$ knows $A_{24}$, and user~$3$
knows $A_{34}$. 
%By lemma \ref{lem:An-achievable-generator}, we can use 
We can use $\frac{3F}{10}-\frac{F}{10}$ random linear combinations of the bits in $\mathcal{O}_{1,4}$. So $\mathcal{C}_{1,4}$ has $\frac{F}{5}$ bits.
By using the same method to encode all the groups, the numbers of bits in $\mathcal{C}_{1,0}$, $\mathcal{C}_{1,4}$, $\mathcal{C}_{1,5}$, $\mathcal{C}_{1,45}$, $\mathcal{C}_{2,0}$, $\mathcal{C}_{2,1}$, $\mathcal{C}_{2,2}$, $\mathcal{C}_{2,3}$, $\mathcal{C}_{2,12}$, $\mathcal{C}_{2,13}$, $\mathcal{C}_{2,23}$ are $\frac{F}{10}$, $\frac{F}{5}$, $\frac{F}{5}$, $\frac{F}{10}$, $0$, $\frac{F}{10}$, $\frac{F}{10}$, $\frac{F}{10}$, $\frac{F}{10}$, $\frac{F}{10}$, $\frac{F}{10}$, respectively. The total number of bits in $\mathcal{C}_\text{step1}$ is  $\frac{6F}{5}$. %therefore

In step~2, it is easy to check that among all the codes $\mathcal{C}_\text{step1}$,
user~$1$ knows $\mathcal{C}_{2,1}$, $\mathcal{C}_{2,12}$ and $\mathcal{C}_{2,13}$, i.e., $\frac{3F}{10}$ bits.  
Similarly, in $\mathcal{C}_\text{step1}$
each user knows $\frac{3F}{10}$ bits.
Hence we can use $\frac{6F}{5}-\frac{3F}{10}=\frac{9F}{10}$ random linear combinations of the bits in $\mathcal{C}_\text{step1}$.
As a result each user can recover each sub-files of his
desired file and the load is $0.9$ while the MNS in~\cite{dvbt2fundamental} requires $1$. This represents $10\%$ saving over the MNS scheme.

\subsection{Proof of Theorem~\ref{thm2 decentralized load}}
\label{sec:newach:decentralized}

Following similar steps to~\cite{decentralizedcoded}, we can extend our proposed delivery method
to decentralized systems as well. Note that since in decentralized systems
no coordination among users is possible, we can not utilize the caching
scheme in~\cite{dvbt2fundamental}.

\subsubsection*{Placement Phase}
The cache placement phase is the same as in~\cite{decentralizedcoded}.
For each $M\in[0,N]$, user $k$ independently caches a subset of
$\frac{MF}{N}$ bits of each file, chosen uniformly at random. Given
the cache content of all the users, we can group the bits of the files
into sets $F_{i,\mathcal{W}}$, where $F_{i,\mathcal{W}}$ is the set of bits 
of file $i$ which are only known by the users in $\mathcal{W}\subseteq[1:K]$. 
By \emph{Law of Large Numbers} we have
$$
\frac{|F_{i,\mathcal{W}}|}{F} \approx \left(\frac{M}{N}\right)^{|\mathcal{W}|}\left(1-\frac{M}{N}\right)^{K-|\mathcal{W}|},\,\,\textrm{ for $F\gg1$.}
$$

\subsubsection*{Delivery Phase}
We divide the sub-files into groups, $DG_{i}=\{F_{i,\mathcal{W}}:|\mathcal{W}|=i\}$
where $i\in[0:K-1]$. The delivery phase described for centralized systems can be used for
the sub-files of $DG_{i}$ for each $i$. 

If we transmit all the coded bits of the groups,
each user can recover his desired file. The %worst-case 
load of the proposed method for decentralized systems is thus 
\begin{align*}
 & R_\text{d}(M)
 =\sum\limits _{i=0}^{\left\lfloor t_\text{th}\right\rfloor }\left(N\binom{K-1}{i}-(N-1)\binom{K-2}{i-1}\right) \cdot \nonumber \\
 & q^{i}\left(1-q\right)^{K-i}+\sum\limits _{i=\left\lfloor t_\text{th}\right\rfloor +1}^{K-1}\binom{K}{i+1}q^{i}\left(1-q\right)^{K-i},
\end{align*}
where $t_\text{th}:=KM_\text{th}/N$ and $q:=\frac{M}{N}$.  
After some simple algebraic manipulations, it is easy to check $R_\text{d}(M)$ can be expressed as in~\eqref{eq:CDF forme rd(M)}.
Finally,
the memory-load trade-off of the proposed scheme is the lower
convex envelope of $R_\text{d}(M)$. \hfill\qed

\subsection{Sketch of the Proof of Theorem~\ref{thm3 uncoded placement optimality}}
\label{sec:newach:optimality}
Assume each file is demanded by at least one user. We denote the worst-case load under the constraint of uncoded placement by $R_\text{u}(M)$.
We choose $N$ users with different demands in the user set $[1:K]$. The chosen user set
is denoted by $\mathcal{C}=\{c_{1},c_{2},...,c_{N}\}$ where $c_{1}<c_{2}<...<c_{N}$
and $c_{i}\in[1:K]$. We assume user $c_{i}$ demands $d_{c_{i}}$,
where $d_{c_{i}},i\in[1:N]$ and $d_{c_{i}}\neq d_{c_{j}}$ if $i\neq j$.
By considering uncoded placement and that other users do not require any file, the delivery phase is
an index coding problem where each message is demanded by only one
user. We denote the worst-case load of the above case by $n(M)$. It is obvious that $R_\text{u}(M)\geq n(M)$. Hence we can use the same method as~\cite{ontheoptimality} based
on the index coding graph where each node represents a sub-file demanded
by one user as argued in~\cite{ontheoptimality}. The only difference is that $\mathbf{u}=(u_{1},u_{2},...,u_{N})$ is a permutation of $\mathcal{C}$. So by following~\cite{ontheoptimality}, it is not difficult to generate the following outer bound for $n(M)$,
\begin{align}
n(M)\geq\sum\limits _{i=0}^{K}\frac{\binom{K-1}{i}+\binom{K-2}{i}+...+\binom{K-N}{i}}{N\binom{K}{i}}x_{i},\label{eq:n(m)}
\\
x_{0}+x_{1}+...+x_{K}=N,\label{eq:file size}
\\
x_{1}+2x_{2}+...+ix_{i}+...+Kx_{K}=KM,\label{eq:cache size}
\end{align}
where $x_{t}$ is the total length of the sub-files that are known by $t$ users, $t\in[0:K]$.
For $N=2$, we eliminate $x_{t}$ for $t\in[0:K]$ in the system of inequalities~\eqref{eq:n(m)}-\eqref{eq:cache size} and get an outer bound for the load $n(M)$. In~\cite{ontheoptimality} we proposed an elimination method to this kind of problem. Please find the details of the elimination in Appendix. Finally, we can see that the above outer bound for $n(M)$ coincides with the lower convex envelope $R_\text{co}(M)$ in~\eqref{eq:rpofM1} for $N=2$. Next we give an example to understand the elimination method.

\subsubsection*{Example}
%By continuing the example 
In the Section~\ref{sec:newach:centralized} where $N=2$, $K=5$ and $M=0.8$, it was shown that the proposed delivery scheme leads to a load equal to $0.9$. Now we prove its optimality. %using the above method.

From expressions~\eqref{eq:n(m)}-\eqref{eq:cache size}, we have that 
\begin{align}
n(M)\geq\sum\limits _{i=0}^{4}\frac{(5-i)(8-i)}{40}x_{i},
\label{eq:examplecase2}
\\
x_{0}+x_{1}+x_{2}+x_{3}+x_{4}+x_{5}=2,
\label{eq:examplefilesize}
\\
x_{1}+2x_{2}+3x_{3}+4x_{4}+5x_{5}=5M.
\label{eq:examplecachesize}
\end{align}
Then we sum~\eqref{eq:examplefilesize}$\times\frac{19}{20}$ and~\eqref{eq:examplecachesize}$\times\frac{-1}{4}$, to find
\begin{equation}
-\frac{7}{10}x_{1}-\frac{9}{20}x_{2}-\frac{1}{5}x_{3}+\frac{1}{20}x_{4}+\frac{3}{10}x_{5}+\frac{19}{10}-\frac{5}{4}M=0.\label{eq:sum file cache}
\end{equation}
At last we take~\eqref{eq:sum file cache} into~\eqref{eq:examplecase2}, and we can have
\begin{align*}
n(M) & \geq\frac{19}{10}-\frac{5}{4}M+\frac{1}{20}x_{0}+\frac{1}{20}x_{3}+\frac{3}{20}x_{4}+\frac{3}{10}x_{5}\nonumber \\
 & \geq\frac{19}{10}-\frac{5}{4}M.
\end{align*}
When $M=4/5$, $R_\mathrm{u}(M)\geq n(M)\geq0.9$ which is equal to the load of the proposed scheme. By using the same method we can know that for any $K>N=2$ and $M\in[0,N]$, the proposed scheme is optimal under the constraint of uncoded placement.
%\paragraph*{Example} {\red Studying analytically a decentralized system
%is quite lengthy and involved, thus we refer the reader to the numerical example in the next Section.}

%\section{On the case of $N=2$ files} 
%\label{sec:optN=2}
%With a similar method to the one in~\cite{ontheoptimality}, we can prove the following theorem. The detailed proof can %be found in an extended version of this manuscript~\cite{?inprep?}.
%\begin{thm}
%\label{thm1 uncoded placement} 
%The minimal load of the worst case among all the possible demands under the constraint of uncoded cache placement and %$K>N=2$ for the aforementioned centralized system, is achieved by the proposed scheme including MA\&N's placement phase and the proposed two-step delivery phase with the load in~\eqref{eq:rp of M 1}. 
%\end{thm}
%{\red add something, even if this is only to make a ref to a result in an extended version of the paper.}

\section{Numerical Results}
\label{sec:num}
%We also divide this section into two parts, simulations for centralized
%system and for decentralized system respectively. Each curve in the simulation
%representing the memory-load trade-off is is the lower convex envelope
%of discrete points.

\begin{figure}%[tbh]
\centering{}
\includegraphics[scale=0.6]{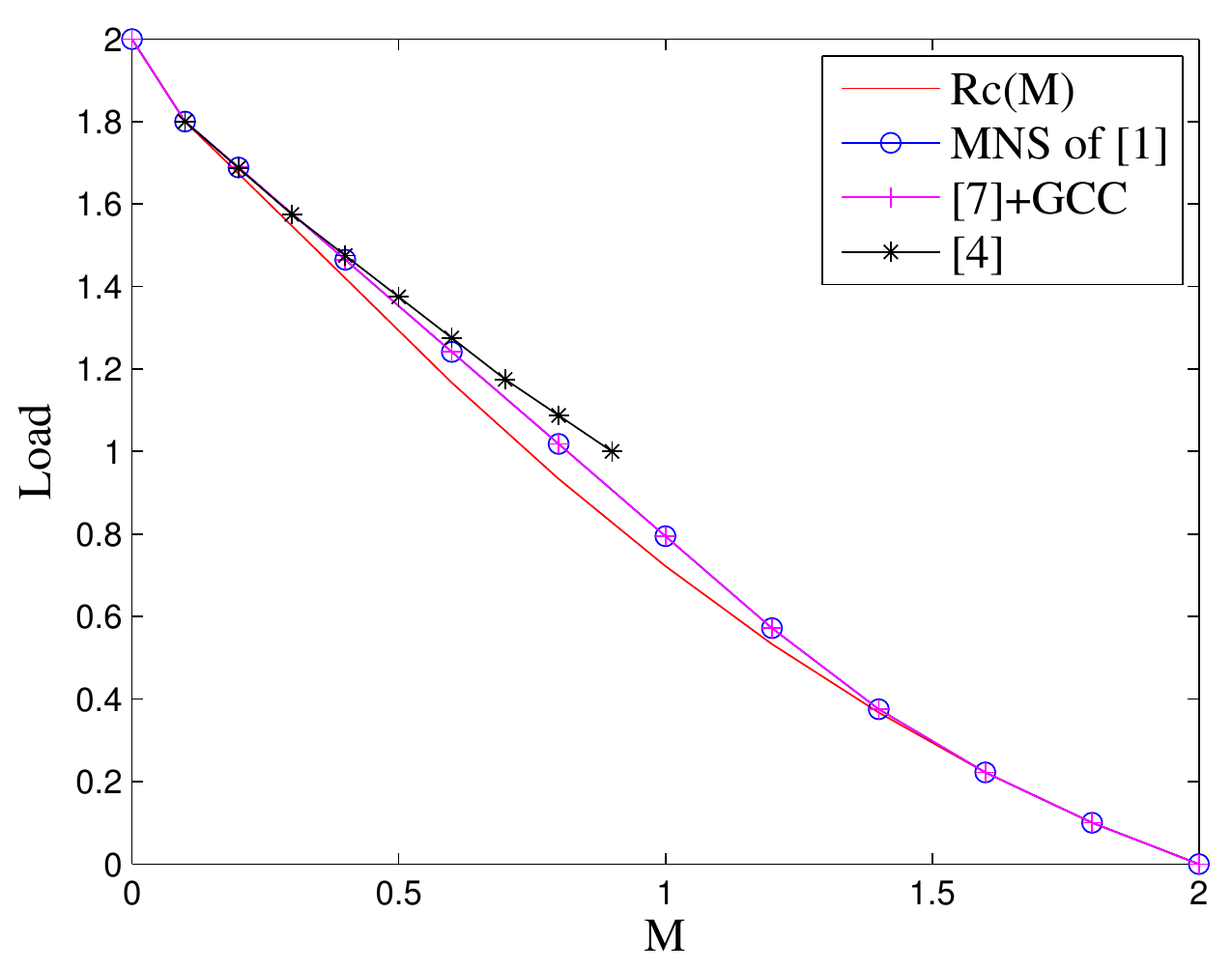}
\caption{The memory-load trade-off for a centralized system with $N=2$ and
$K=10$.}
\label{fig:excen}
\end{figure}

\subsection{Centralized Systems}
%In this system, coordination is allowed in the placement phase. 
We compare the achievable load with our proposed scheme in~\eqref{eq:Rp(M)}
with that of the schemes in~\cite{dvbt2fundamental,kuserstwofiles,groupcastindexcoding}.
Since  the scheme in~\cite{smallbufferusers}
is optimal when $0\leq M\leq\frac{1}{K}$, 
we memory-share each considered scheme with the one in~\cite{smallbufferusers}. 
Note that~\cite{groupcastindexcoding} uses the local chromatic number, whose computation is NP-hard;
here in order to simplify the computations we use the approximate algorithms
GCC, HgLC and GRASP proposed in~\cite{simplifiedgroupeindexcoding,grasp}. Numerically we find that for centralized system GCC performs better than the other simplification methods. Therefore, in order to have a less cluttered figure, we only plot GCC. We also do the numerical evaluations for the MNS and the scheme in~\cite{kuserstwofiles}. Fig.~\ref{fig:excen} shows  the memory-load trade-offs for a centralized
system with $N=2$ and $K=10$. We can see the scheme in~\cite{groupcastindexcoding} with GCC has the same performance as the MNS and the improvement from the scheme in~\cite{kuserstwofiles} to the MNS is negligible. Furthermore, the proposed scheme improves on the MNS. For instance, when $M=1$, the proposed scheme (with load $0.722$) reduces $9.1\%$ of the load of the MNS ($0.794$).

\subsection{Decentralized Systems}
In decentralized scenarios, the caching schemes in~\cite{smallbufferusers}
and~\cite{kuserstwofiles} with coordinated cache placement can not
be used. The scheme in~\cite{detailsgroupe} is similar to the one
in~\cite{groupcastindexcoding}, where the main difference relies on the use of random placement for the first one. In order to compute the local chromatic
number, the approximate GCC, HgLC and GRASP algorithms are used. The authors in~\cite{simplifiedgroupeindexcoding,grasp} claimed that for decentralized system with uniform demands and infinite file size, GCC performs better than other simplification methods. Hence, we compare the scheme in~\cite{groupcastindexcoding} with GCC, $R_\text{d}(M)$ in~\eqref{eq:CDF forme rd(M)} and
the decentralized MNS in~\cite{decentralizedcoded} in the numerical evaluations.

\begin{figure}%[tbh]
\centering{}
\includegraphics[scale=0.6]{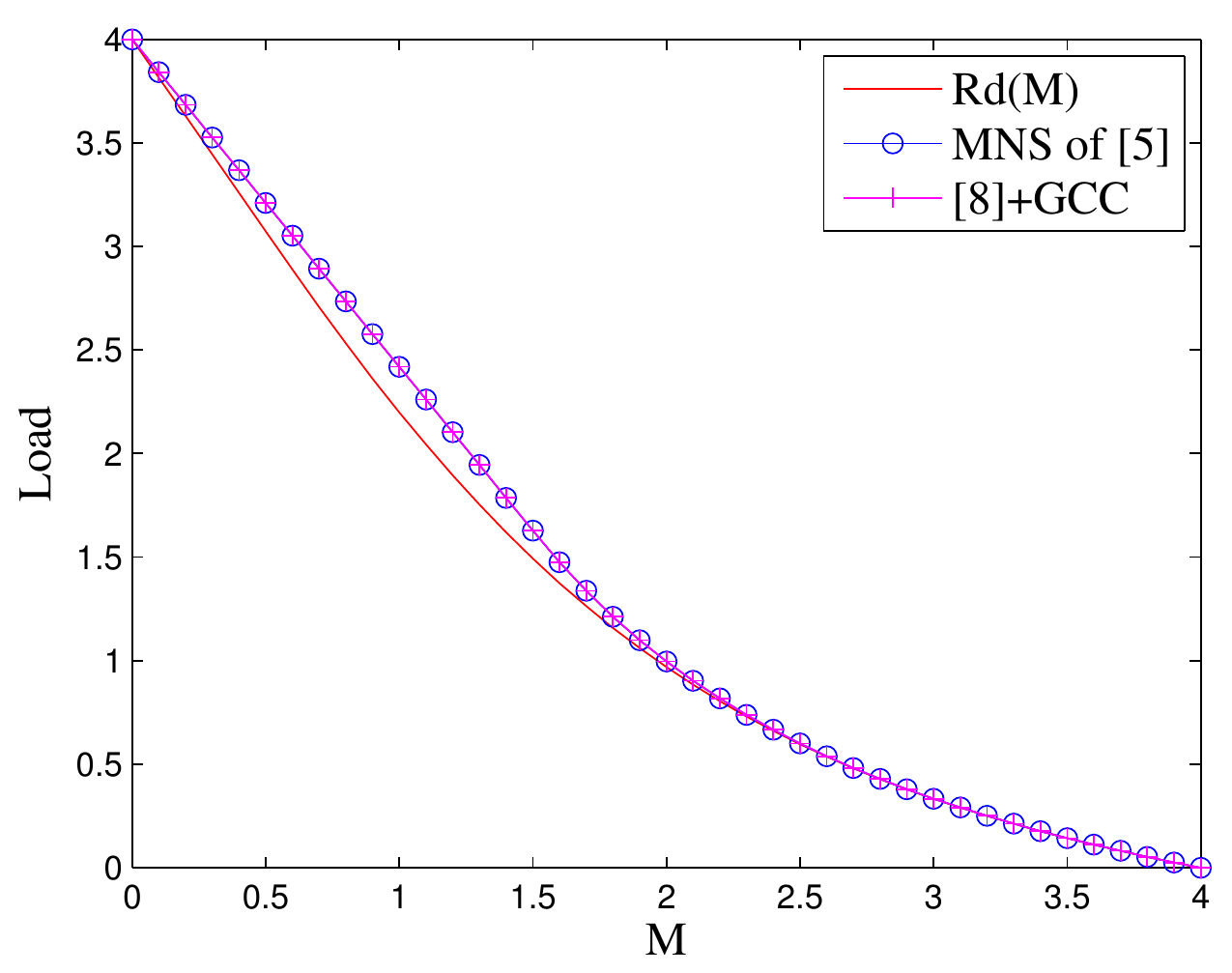}
\caption{The memory-load trade-off for a decentralized system with $N=4$ and
$K=8$.}
\label{fig:exdec}
\end{figure}

Fig.~\ref{fig:exdec} shows  the memory-load trade-offs for a decentralized
system with $N=4$ and $K=8$. The decentralized MNS and the scheme in~\cite{groupcastindexcoding} with GCC have the same performance, while the proposed scheme performs better than the other ones. For instance, when $M=1.2$, the proposed scheme (with load $1.894$) reduces $9.9\%$ of the load of the decentralized MNS ($2.102$).

\section{Conclusion and Further Work}
\label{sec:conclusions}
We investigated the caching
problem for centralized and decentralized systems with more users than files, 
%of~\cite{dvbt2fundamental} and~\cite{decentralizedcoded} respectively, i.e.,$N<K$, 
which implies a file may be demanded by several users.
%With the same cache placements as~\cite{dvbt2fundamental} and~\cite{decentralizedcoded},
We proposed a novel delivery method leveraging multicasting opportunities with the cache placements of the MNS. We showed that under the constraint of uncoded placement and $K>N=2$, the proposed scheme is optimal for centralized systems. Furthermore, numerical results showed that our proposed scheme outperforms previous schemes for both centralized and decentralized systems.
%which is shown to be better than the ones in~\cite{dvbt2fundamental} and~\cite{decentralizedcoded} for centralized system and decentralized system respectively.  

Further work includes studying coded cache placement and coded delivery schemes while establishing outer bounds and optimality results beyond those derived in this paper.
%will be in two directions. In the simulation we find for centralized systems, the gain from our proposed method over the MA\&N scheme  combining the scheme in~\cite{smallbufferusers}, becomes small when $N$ grows. We want to modify our scheme to improve that gain. On the other hand, we want to design a scheme with different placement phase to have a further approach to the outer bound. 

\section*{Acknowledgments}
The work of K. Wan and D. Tuninetti is supported by Labex DigiCosme and in part by NSF~1527059, respectively. 

\appendix
%\begin{IEEEproof}
Firstly we recall the \emph{Vandermonde's} identity:
\begin{align*}
\binom{m+n}{r}=\sum\limits _{k=0}^{r}\binom{m}{k}\binom{n}{r-k}.
\end{align*}

From~\eqref{eq:app 1},
\begin{align*}
 & \sum_{\mathcal{J}\subseteq[1:K]\setminus\mathcal{G}_{i}:\textrm{max}\{0,t-\mathcal{G}_{i}+1\}\leq |\mathcal{J}|\leq t}\binom{|\mathcal{G}_{i}|-1}{t-|\mathcal{J}|}\\
 & =\sum\limits _{k=\textrm{max}\{0,t-\mathcal{G}_{i}+1\}}^{\mathrm{min}(|[1:K]\setminus\mathcal{G}_{i}|,t)}\sum_{\mathcal{J}\subseteq[1:K]\setminus\mathcal{G}_{i}:|\mathcal{J}|=k}\binom{|\mathcal{G}_{i}|-1}{t-k}\\
 & =\sum\limits _{k=\textrm{max}\{0,t-\mathcal{G}_{i}+1\}}^{\mathrm{min}(|[1:K]\setminus\mathcal{G}_{i}|,t)}\binom{|\mathcal{G}_{i}|-1}{t-k}\binom{K-|\mathcal{G}_{i}|}{k}=\binom{K-1}{t}.
\end{align*}

Similarly from~\eqref{eq:app2}, 
\begin{align*}
 & \sum_{\mathcal{J}\subseteq[1:K]\setminus\mathcal{G}_{i}:\textrm{max}\{0,t-\mathcal{G}_{i}+1\}\leq |\mathcal{J}|\leq t,j\in\mathcal{J}}\binom{|\mathcal{G}_{i}|-1}{t-|\mathcal{J}|}\\
 & =\sum\limits _{k=\textrm{max}\{0,t-\mathcal{G}_{i}+1\}}^{\mathrm{min}(|[1:K]\setminus\mathcal{G}_{i}|,t)}\sum_{\mathcal{J}\subseteq[1:K]\setminus\mathcal{G}_{i}:|\mathcal{J}|\leq k,j\in\mathcal{J}}\binom{|\mathcal{G}_{i}|-1}{t-k}\\
 & =\sum\limits _{k=\textrm{max}\{1,t-\mathcal{G}_{i}+1\}}^{\mathrm{min}(|[1:K]\setminus\mathcal{G}_{i}|,t)}\binom{|\mathcal{G}_{i}|-1}{t-k}\binom{K-|\mathcal{G}_{i}|-1}{k-1}=\binom{K-2}{t-1}.
\end{align*}

Finally we will show the elimination of $x_{t}$ for $t\in[0:K]$ in the system of inequalities~\eqref{eq:n(m)}-\eqref{eq:cache size}.

If $N=2$,~\eqref{eq:n(m)}-\eqref{eq:cache size} becomes
\begin{align}
n(M)\geq\sum\limits _{i=0}^{K}\frac{(K-i)(2K-i-2)}{2K(K-1)}x_{i},
\label{eq:n(m)for2n}
\\
x_{0}+x_{1}+...+x_{K}=2,
\label{eq:N2file size}
\\
x_{1}+2x_{2}+...+ix_{i}+...+Kx_{K}=KM.
\label{eq:N2cache size}
\end{align}

For a $q\in[1:K]$ we want to eliminate $x_{q}$ and $x_{q-1}$
in~\eqref{eq:n(m)for2n} by the help of~\eqref{eq:N2file size} and~\eqref{eq:N2cache size}.

From~\eqref{eq:N2file size}, we have
\begin{align}
 & \frac{2K^{2}-2K-q^{2}+q}{2K(K-1)}(x_{q-1}+x_{q})\nonumber \\
 & =\frac{2K^{2}-2K-q^{2}+q}{2K(K-1)}(2-\sum_{i\in[0:K]:i\neq q-1,q}x_{i}).\label{eq:timesfilesize}
\end{align}

From~\eqref{eq:N2cache size}, we have
\begin{align}
 & \frac{2q-3K+1}{2K(K-1)}(q-1)x_{q-1}+\frac{2q-3K+1}{2K(K-1)}qx_{q}\nonumber \\
 & =\frac{2q-3K+1}{2K(K-1)}KM-\frac{2q-3K+1}{2K(K-1)}\sum_{i\in[0:K]:i\neq q-1,q}ix_{i}.\label{eq:timescachesize}
\end{align}

Then we sum~\eqref{eq:timesfilesize} and~\eqref{eq:timescachesize},
\begin{align}
 & \frac{(K-q)(2K-q-2)}{2K(K-1)}x_{q-1}+\frac{(K-q+1)(2K-q-1)}{2K(K-1)}x_{q}\nonumber \\
 & =\frac{2K^{2}-2K-q^{2}+q}{2K(K-1)}(2-\sum_{i\in[0:K]:i\neq q-1,q}x_{i})+\nonumber \\
 & \frac{2q-3K+1}{2(K-1)}M+\frac{2q-3K+1}{2K(K-1)}\sum_{i\in[0:K]:i\neq q-1,q}ix_{i}\nonumber \\
 & =\frac{2q-3K+1}{2(K-1)}M+\frac{2K^{2}-2K-q^{2}+q}{K(K-1)}+\nonumber \\
 & \sum_{i\in[0:K]:i\neq q-1,q}\frac{2K^{2}+2K+3Ki+q^{2}-q-2qi-i}{2K(K-1)}x_{i}.\label{eq:takesum}
\end{align}

Take~\eqref{eq:takesum} into~\eqref{eq:n(m)for2n},
\begin{align}
n(M) & \geq\sum\limits _{i=0}^{K}\frac{(K-i)(2K-i-2)}{2K(K-1)}x_{i}\nonumber \\
 & \geq\frac{2K^{2}-2K-q^{2}+q}{K(K-1)}+\frac{2q-3K+1}{2(K-1)}M\nonumber \\
 & +\sum\limits _{i=0}^{K}\frac{(q-i)(q-i-1)}{2K(K-1)}x_{i}\nonumber \\
 & \geq\frac{2K^{2}-2K-q^{2}+q}{K(K-1)}+\frac{2q-3K+1}{2(K-1)}M.\label{eq:result}
\end{align}

When $M=Nq/K$,~\eqref{eq:result} becomes
\begin{align*}
n(M) & \geq\frac{2K^{2}-2K-q^{2}+q}{K(K-1)}+\frac{2q-3K+1}{2(K-1)}\frac{2q}{K}\\
 & =\frac{2(K-q)}{K}-\frac{q(K-q)}{K(K-1)}.
\end{align*}

When $M=N(q-1)/K$,~\eqref{eq:result} becomes
\begin{align*}
n(M) & \geq\frac{2K^{2}-2K-q^{2}+q}{K(K-1)}+\frac{2q-3K+1}{2(K-1)}\frac{2(q-1)}{K}\\
 & =\frac{2(K-q+1)}{K}-\frac{(q-1)(K-q+1)}{K(K-1)}.
\end{align*}

Hence for $\frac{N(q-1)}{K}\leq M\leq\frac{Nq}{K}$, we can see the
linear outer bound of $n(M)$ in~\eqref{eq:result}, as well as $R_\text{u}(M)$,
coincides with the lower convex envelop of the load of our proposed
load in~\eqref{eq:rpofM1}.

%\end{IEEEproof}

\bibliographystyle{IEEEtran}
\bibliography{IEEEabrv,IEEEexample}

\end{document}